\documentclass[aps,reprint,amsmath,amssymb,showpacs,superscriptaddress]{revtex4-2}

\usepackage{graphicx,color,epsfig}
\usepackage{dcolumn}
\usepackage{bm}
\usepackage[caption=false]{subfig}
\usepackage{color}
\usepackage[normalem]{ulem}
\usepackage[T1]{fontenc}
\usepackage[utf8x]{inputenc}
\usepackage[normalem]{ulem}

\begin{document}
\title{Interplay of Reward and Size of Groups in the Optional Public Goods Game}

\author{Eduardo V. Stock}
\affiliation{Instituto de Física, Universidade Federal do Rio Grande
  do Sul, Caixa Postal 15051, 90501-970 Porto Alegre RS, Brazil}
\email{eduardo.stock@ufrgs.br,sgonc@if.ufrgs.br,rdasilva@if.ufrgs.br}

\author{Pablo Valverde}
\affiliation{Pontificia Universidad Católica del Ecuador, Facultad de Ciencias Exactas y Naturales, Quito 170525, Ecuador}
\email{pblvalverde@gmail.com}

\author{Juan Carlos González-Avella}
\affiliation{Institute for Cross-Disciplinary Physics and Complex Systems, UIB-CSIC, Palma de Mallorca 07122, Spain; Advanced Programming Solutions SL, Palma de Mallorca 07120, Spain}
\email{avellaj@gmail.com}

\author{José Roberto Iglesias}
\affiliation{Instituto de Física, Universidade Federal do Rio Grande
  do Sul, Caixa Postal 15051, 90501-970 Porto Alegre RS, Brazil}
  \affiliation{Instituto Nacional de Ciência e Tecnologia de Sistemas Complexos, CBPF, Rio de Janeiro 22290-180, RJ, Brazil}
\email{iglesias@if.ufrgs.br}

\author{Sebastián Gonçalves}
\affiliation{Instituto de Física, Universidade Federal do Rio Grande
  do Sul, Caixa Postal 15051, 90501-970 Porto Alegre RS, Brazil}
\email{sgonc@if.ufrgs.br}

\author{Roberto da Silva}
\affiliation{Instituto de Física, Universidade Federal do Rio Grande
  do Sul, Caixa Postal 15051, 90501-970 Porto Alegre RS, Brazil}
\email{rdasilva@if.ufrgs.br}

\date{\today}

\begin{abstract}
The Optional Public Goods Game is a three-strategy game in which an
individual can play as a cooperator or defector or decide not to
participate.  
Despite its simplicity, this model can effectively represent many human social dilemmas, such as those found in the use of public services, environmental concerns, or other activities related to society. In this contribution, we present a comprehensive analysis of the conditions under which spontaneous, sustained cooperation emerges and the characteristics of these cooperative states.
Through simulations, we demonstrate the conditions leading to the coexistence of the three strategies in a steady state or the alternate dominance of each strategy in a rock-paper-scissors fashion.
The results identify each of the possible scenarios in terms of two key parameters: the multiplication rate of the public good game (reward) and the size of the group of potential players. 
We also discuss other details of the game that may influence the appearance of cycles, along with relevant characteristics of these cycles, such as the prevalence of cooperation.
\end{abstract}

\maketitle

\section{Introduction}

The emergence of cooperative behavior among independent individuals remains a central challenge in both biological and social systems~\cite{SMITH198643}. The almost inevitable phenomenon of people profiting from a social endeavor without the burden of contribution has been a major focus of theoretical and experimental studies. 
Similarly, scientists are trying to understand the factors that favor cooperation~\cite{Grujic2012,Granulo2023}. While punishment~\cite {Fehr2002,botta2021,Li2021} or reward mechanisms~\cite{Forsyth2011} can promote cooperation, they depend on the ability to identify individuals. In this sense, a range of different mechanisms have been proposed in different topologies to understand these dynamics. This is the case of information sharing between players~\cite{Li2024}, conditional cooperation~\cite{Andreozzi2020}, conditional dissociation of defectors~\cite{Krivan2020}, imposing a minimal number of cooperators~\cite{DeJaegher2020}, and others in attempts to prevent defection. 

In particular, Hauert et al. showed that voluntary participation in public goods games (PGGs) introduces abstention as a stabilizing mechanism, allowing cooperators and defectors to coexist~\cite{Hauert2002}. 
Similarly, Ma et al. demonstrated that cycles of cooperation and free-riding can emerge in social systems, with free-riders and cooperators varying periodically depending on community size and individual costs of cooperation and defection~\cite{Ma2009}.
From a game theory perspective, a public goods game (PGG) extends the classic prisoner's dilemma, involving two or more players who can either cooperate or defect in contributing to a shared public good. 

Civilini et al. have shown that higher-order interactions between players in complex networks can lead to explosive cooperation, unlike traditional pairwise interactions \cite{PhysRevLett.132.167401}. Similarly, Andreozzi et al. \cite{Andreozzi2020} discovered that conditional cooperators are more likely to defect unconditionally in an anonymous social dilemma experiment, where multiple opportunities to switch strategies are available before each game, rather than just once \cite{Fischbacher2010}. Furthermore, agent-based models suggest that, when sufficient noise is present in the dynamics, introducing a few altruistic players can substantially boost and sustain cooperation levels within the population \cite{Battu2023}.

In this context, the PGG provides an insightful framework for investigating public spending on community projects such as roads, bridges, or libraries. Players are given the opportunity to invest their money in a shared fund, with profits ---generated from sources such as tolls or membership fees--- distributed equally among all participants, regardless of individual contributions. 
Ideally, it might seem ``fair'' for individuals with similar financial means to contribute equally to these projects. However, people differ in their social and financial circumstances, meaning that some can afford to invest more than others. Since each player is unaware of the contributions of others, a purely ``rational'' player might choose to invest nothing, thus avoiding their responsibility. As a result, for purely rational participants, the dominant strategy is to defect by contributing nothing.

Considering the different approaches to the PGG, an early one involves studies of a single pool where players interact and contribute to the same public good, following dynamics driven by motivation, as modeled by some authors \cite{Ashlock86214,DASILVA2006610,daSilva2008,Silva20102}. This approach became widely known in the context of evolutionary computing as the \textit{public investment game} (see, for example, \cite{Ashlock86214}).

A second approach envisions agents playing in different groups or separate pools, which is the method most commonly explored in most of the contributions to PGG in the literature.

In this context, the alternative Optional Public Goods Game (OPGG) serves as a valuable model for understanding social conflicts and activities where cooperation is essential, such as the provision of public services and environmentally related behaviors. In this version of the PGG, players can choose to abstain from participating in the game, receive a fixed payment, and become the so-called \textit{loners}.

The coexistence of the three strategies: cooperators, defectors, and loners, as well as the cyclical dominance among them ({\em i.e.}, rock-paper-scissors (RPS) dynamics), has been identified as a possible outcome of the OPGG when analyzed using mean-field replicator dynamics~\cite{HAUERT2002187,hauert2003}.

These findings are significant, as both equilibrium and cyclic dominance scenarios suggest spontaneous emergence of cooperation. Replicator dynamics implicitly assume that, in an infinite population, player groups can vary in size, games are played synchronously, and the players have complete knowledge of the payoff of the population when making decisions.

However, these results were demonstrated for specific parameter ranges, and a comprehensive analysis across a broader set of parameters that promote cooperation has yet to be conducted and warrants further investigation. Questions regarding the statistics of the number, size, and duration of cycles are essential to get a complete understanding of the problem. Therefore, an alternative numerical study of this model is necessary, along with numerical simulations that consider the relaxation of assumptions used in the mean-field regime.

In this contribution, we present a comprehensive analysis through simulations of the conditions for the emergence of spontaneous cooperation, either as coexistence in steady equilibrium states or as alternating dominance among strategies in a rock-paper-scissors manner. Our analysis focuses on the simultaneous effects of the multiplication factor ($r$) and the size of the game groups ($S$) on the outcomes of the OPGG. Specifically, we explore how these two key parameters influence the emergence of spontaneous cooperation within communities, whether through stationary coexistence of strategies or cyclic dominance among the three possible strategies.

In Section \ref{model}, we describe the model and the results are presented in Section \ref{simul}. The summaries and conclusions are given in Section \ref{conclu}.

\section{The Model}\label{model}
We consider a population of $N$ agents that can be found in one of three
possible states: C (cooperator), D (defector), or L (loner).  The OPGG
evolves as follows:

\begin{enumerate}
\item Two different agents, here denoted as $i$ and $j$, are randomly chosen from the population.

\item For each of the chosen agents, a set of $S-1$ distinct agents are randomly selected (uniform distribution) from the $N-1$  agents left giving rise to two groups: $S_i$ and $S_j$. Both groups can share one or more agents.

\item Each cooperator in a group contributes to the common pool with a unit of wealth. Defectors participate, but without contribution, while loners stay out of the game getting a fixed payoff rate $\sigma$ proportional to his/her investment $c$ as well.

\item Payoffs are then calculated for the three strategies:
    
\begin{equation} P = \left\{
\begin{array}{l c l} \frac{r N_C}{(N_C+N_D)}-1 & ; & C\\ \\
\frac{r N_C}{(N_C+N_D)} & ; & D\\ \\  \sigma & ; & L\\
\end{array} \right.
\end{equation}

where $r$ is the interest rate of the public good, $N_C$,
$N_D$, and $N_L$ are the number of cooperators, defectors, and loners, respectively, in the local configuration.

\item With complete knowledge of $S_i$ and $S_j$ payoff information, agent $i$ chooses to switch to player $j$ strategy according to a Glauber-dynamics-like transition probability density function
\begin{equation}
P_{i \rightarrow j} = \frac{1}{1 + e^{-(P_j-P_i)/k}}
\label{fermi}
\end{equation} where $k$ is a ``thermal noise'' parameter and measure the level of randomness of the strategy change dynamics, which can be more clearly observed, for example, in \cite{TRAULSEN2007349,Perc_2006}.

\item The steps above are repeated $N$ times to make one Monte Carlo step.

\end{enumerate}

The Glauber dynamics, shown in Eq.\ref{fermi}, was chosen as the transition probability for changing strategies, as in \cite{SZABO200797,Perc_2006}, due to its flexibility in representing various stochastic levels. 
For instance, in the limit $k \rightarrow 0$ (except when $P_i = P_j$), the system approaches a nearly deterministic regime where players are almost certain to change or maintain their strategy based on a higher or lower payoff compared to another player. Conversely, as $k \rightarrow \infty$, the system exhibits purely stochastic behavior, where players have a $1/2$ probability of changing or retaining their strategy.

To investigate the collective dynamical behavior of the system, we follow the state of the system by measuring the
fraction (or density) of individuals in the three possible strategies, $\rho^l_C \equiv \frac{N^l_C}{N}, \rho^l_D \equiv \frac{N^l_D}{N}$ and
$\rho^l_L \equiv \frac{N^l_L}{N}$, cooperators, defectors, and loners, respectively, at a given time step $l$.

Under certain specific circumstances, the system can present a dynamic state where the prevalent strategy changes in a seemingly steady cyclic fashion, called rock-paper-scissor cycles (RPS), which in our case Cooperator-Defector-Loner(CDL) would be a more suitable acronym (or DLC, or CLD). Here, we will denote the number of complete cycles within a specified time frame by $\lambda$. Another important variable to measure the occurrence of RPS cycles was proposed in \cite{VALVERDE201761} and it is defined as
\begin{equation}
    \alpha \equiv \frac{1}{t_{f}}\sum\limits_{j}t_j,
    \label{alpha_def}
\end{equation}
which measures the time fraction of the $\lambda$ RPS cycles within the evolution time frame $t_f$. 

To understand how cycles are measured, we should realize that a given strategy $X$ is generally expected to be prevalent at each time step, that is, $\rho_X^l>\rho^l_{Y,Z}$ (the density of agents using strategy $X$ is greater than the density of agents of the other two strategies at time step $l$). Thus, in a given time interval of evolution, we can have the dominant strategies presenting the following results: $$\cdots \mbox{D-CDDLLLLL-CDDLLLL-CDLLL-CDL-CCCL}\cdots,$$ which corresponds to $\lambda=4$ and $\alpha=(8+7+5+3)/28\approx0.82$.

In the next section, we present the main results obtained via Monte Carlo simulations. For clarity, in this work, we only studied initial conditions where all strategies were equally likely to happen, i.e. $\rho_c^0$=$\rho_d^0$=$\rho_l^0$=1/3. We also considered the loner strategy interest rate to have a fixed value of $\sigma=1$, so there is no further reference to it in this work. 

\section{Numerical Simulations}\label{simul}

\begin{figure*}
\begin{center}
\includegraphics[width=0.48\textwidth]{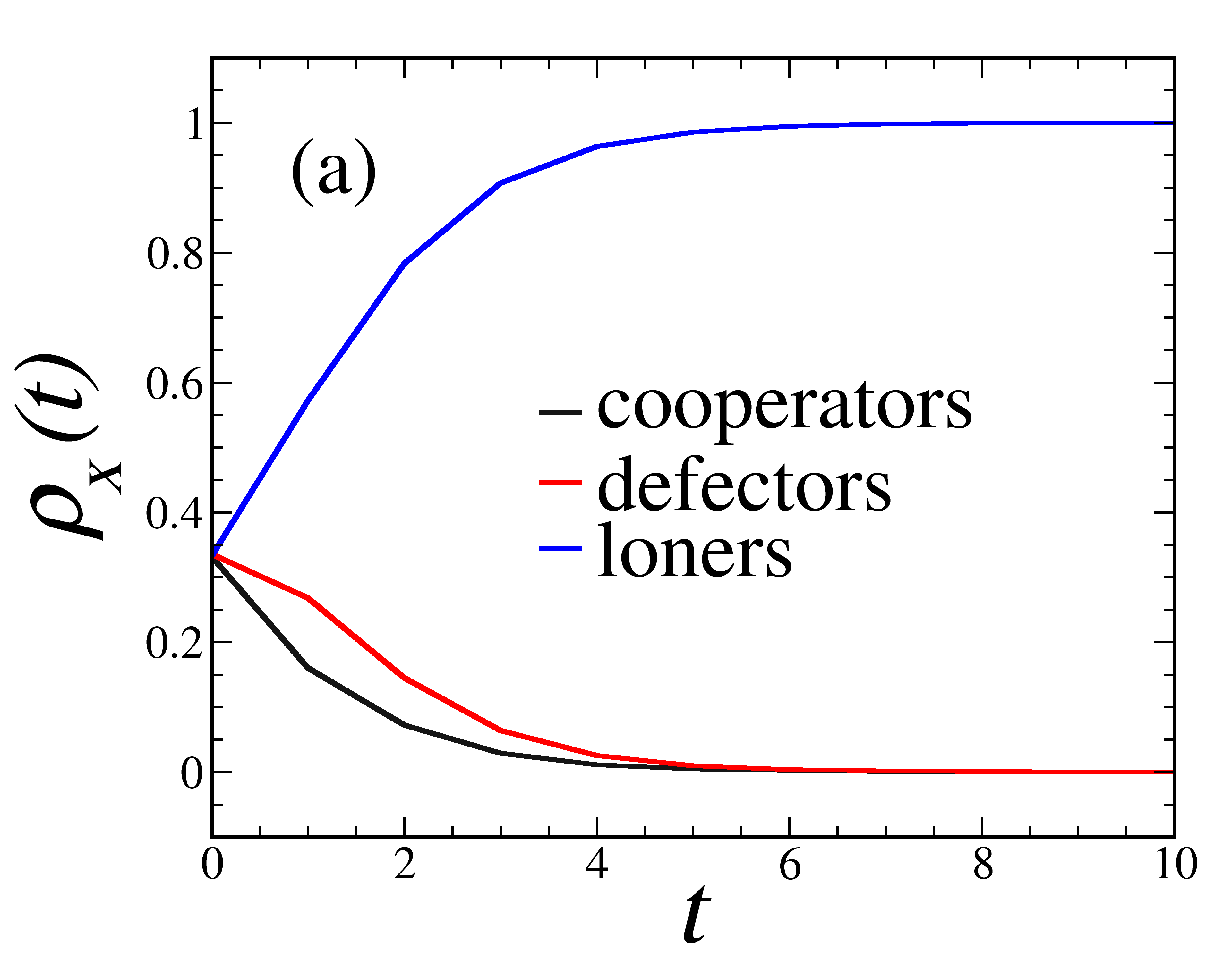}\includegraphics[width=0.48\textwidth]{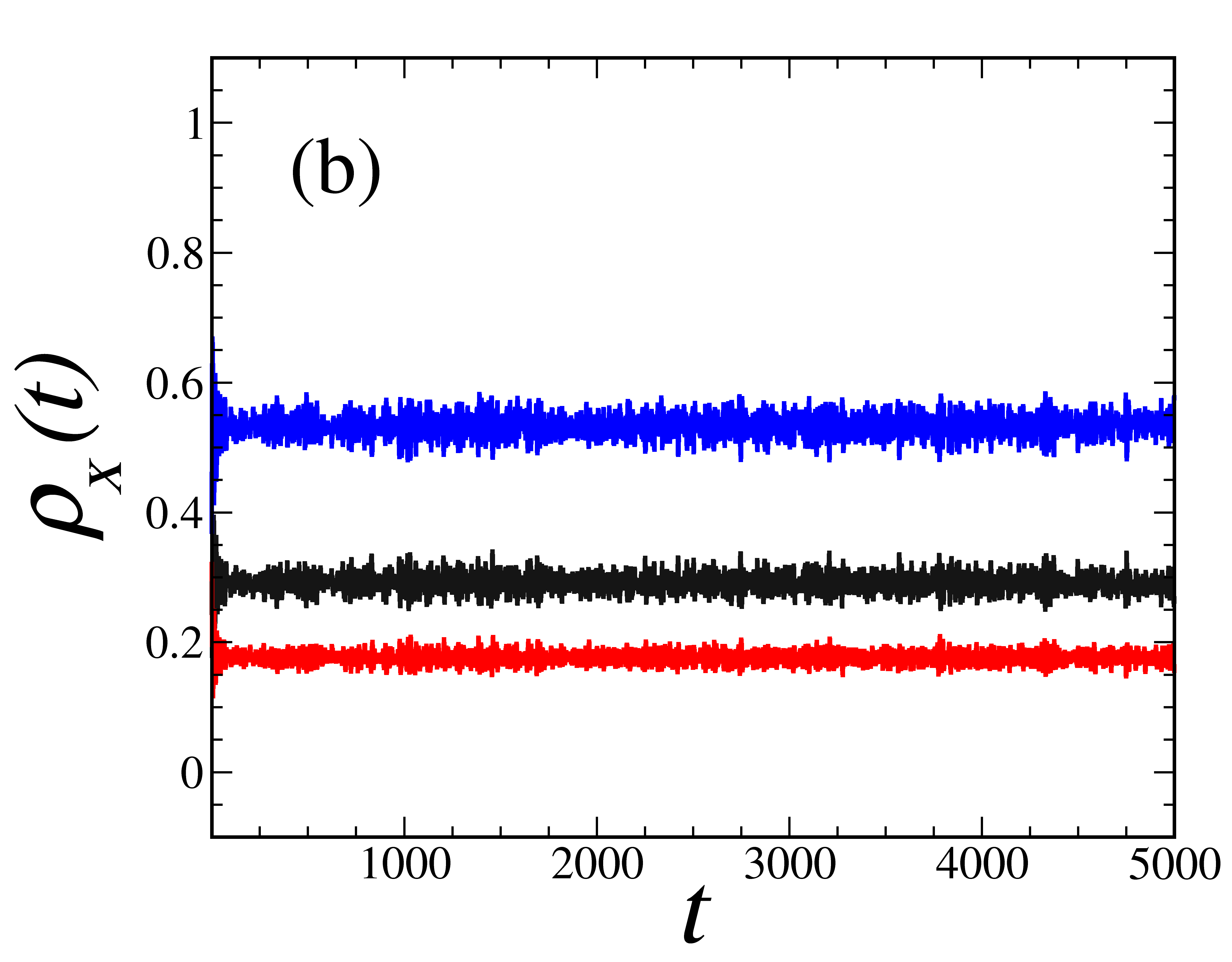}
\includegraphics[width=0.48\textwidth]{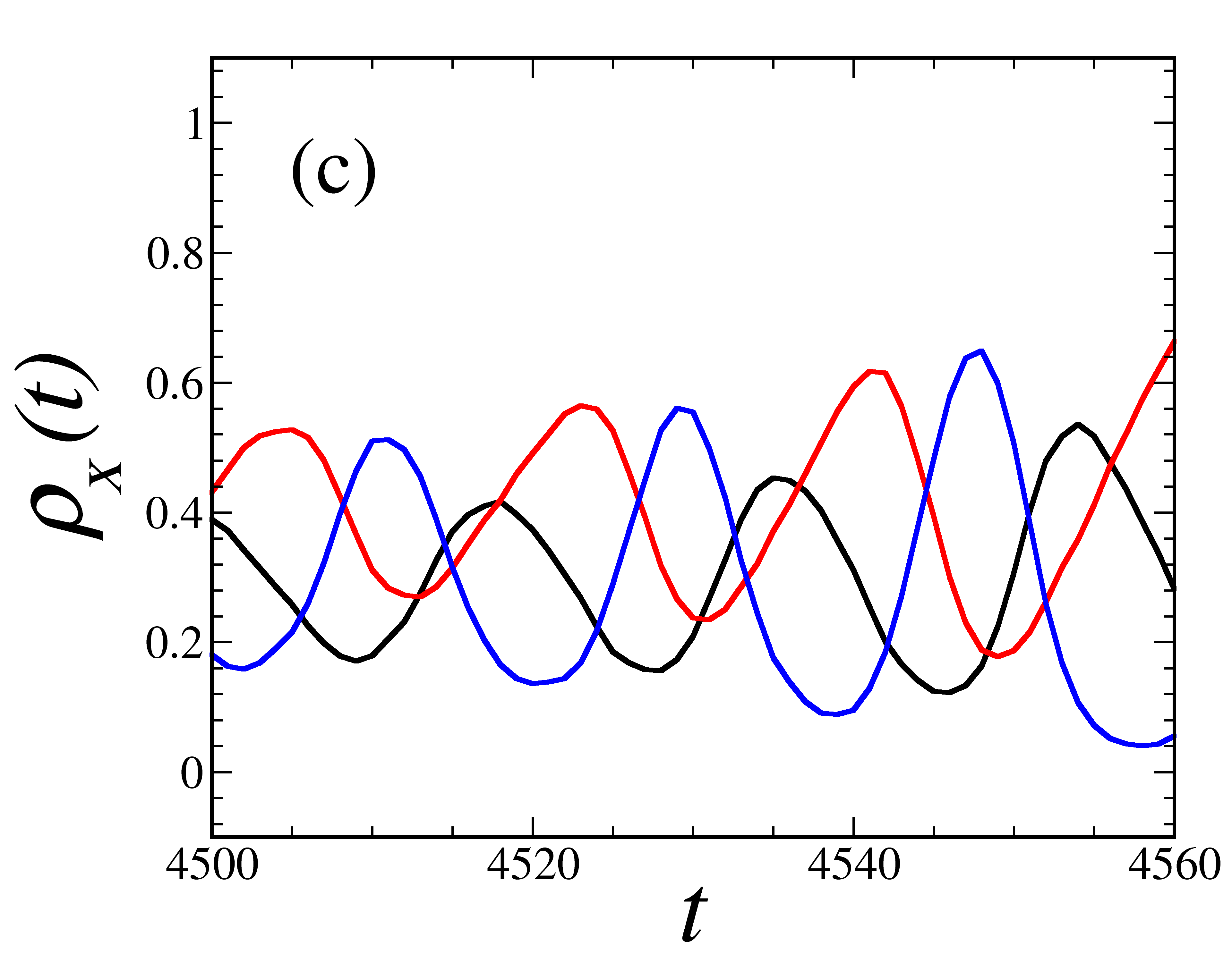}\includegraphics[width=0.48\textwidth]{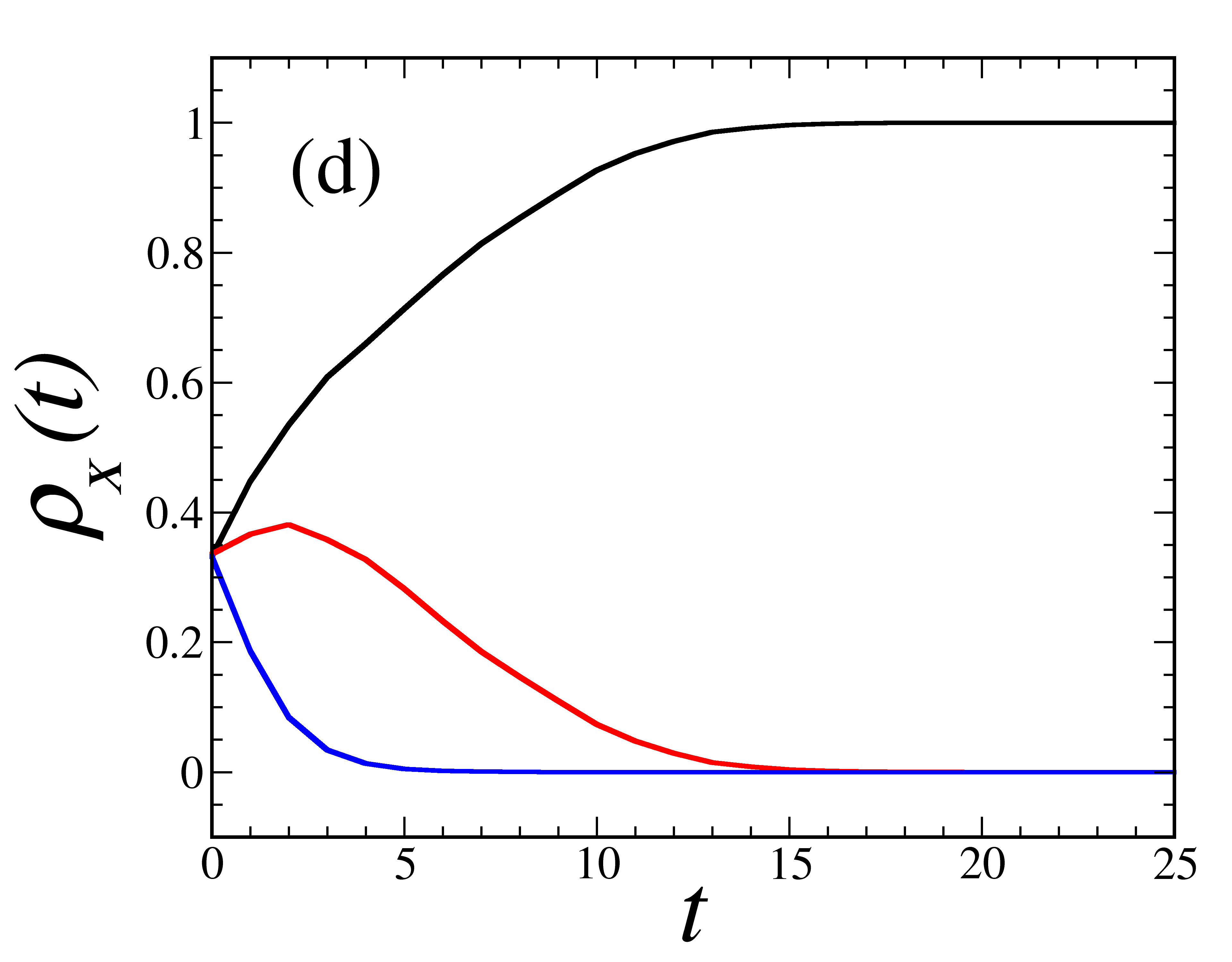}
\end{center}
\caption{Time series of the fraction of the three strategies for (a) $r=1.0$, (b) $r=2.5$, (c) $r=3.5$, and (d) $r=6.0$. Each time series was obtained using $N=10^4$, $S=5$, and $k=0.1$.} \label{fig:time_series}
\end{figure*}

We now show the evolution of the fraction of cooperators, defectors, and loners for different values of $r$ while keeping the parameters fixed at $S=5$ and $k=0.1$. For $r=1.0$, Fig.~\ref{fig:time_series}(a) the dynamics reveal that after a few MC steps, the system reaches an absorbing state where all agents have adopted the loner strategy ($\rho_C=\rho_D=0$ and $\rho_L=1$), which is a quite expected result considering the minimum return for those invited to participate in the public goods game.  In contrast, we show in Fig. ~\ref{fig:time_series}(b) that for $r = 2.5$, the evolution reaches a reasonable steady state with strategies coexisting in a regime where each density presents small fluctuations around its respective mean values.

For $r = 3.5$, there is still a coexistence of strategies, but the collective behavior of the system changes drastically, as we can see in Fig. ~\ref{fig:time_series}(c). The system now presents large oscillations and the prevalent strategy changes in a cyclic fashion corresponding to the emergence of RPS cycles. Finally, in Fig. ~\ref{fig:time_series} we show that for a sufficiently large public goods game return, i.e. $r=6.0$, the system presents a rapid relaxation to an absorbing state where all players cooperate.
\begin{figure}[htbp!]
\begin{center}
\includegraphics[width=1.0\linewidth,angle=0]{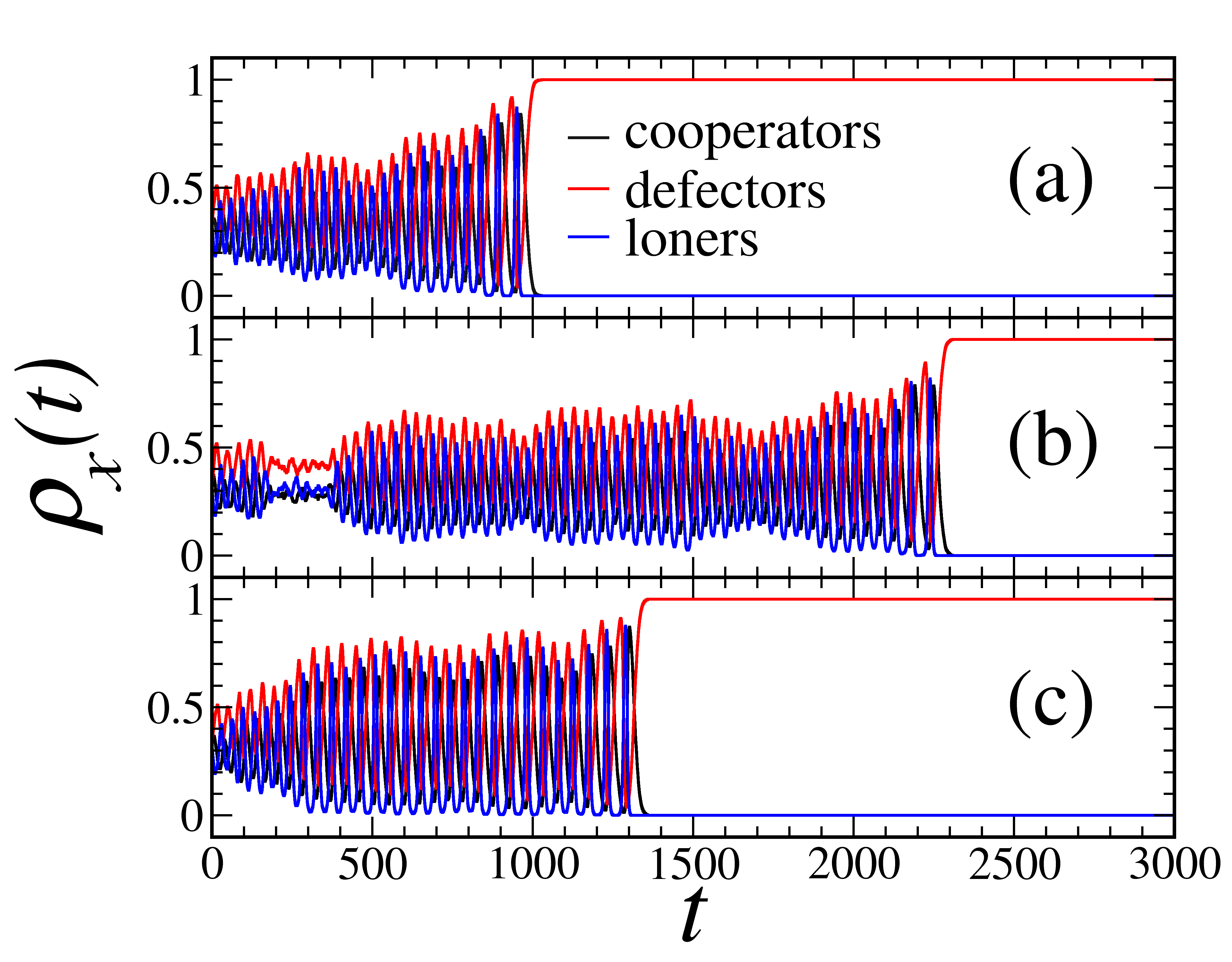}
\end{center}
\caption{Evolution of the strategies densities of three runs obtained with different initial seeds whilst keeping the same parameters $N=10^4$, $S=5$, and $k=1$.}
\label{fig:time_series_seeds}
\end{figure}

The presence of RPS cycles shown in Fig. \ref{fig:time_series}(c) for the parameters specified in the caption is in great agreement with the results obtained by Hauert et al. \cite{HAUERT2002187}. Thus, we investigate a little further the appearance of RPS cycles by showing three different runs (different seeds) in Fig.~\ref{fig:time_series_seeds} obtained with parameters $N=10^4$, $r=3.5$, and $k=1$. It is possible to observe that all three runs present the emergence of RPS cycles with considerably large oscillations until an absorbing state of only defectors is reached. Each run reaches this absorbing state in a different relaxation time after which no changes in the system are observed.

In practical terms, this relaxation time, denoted here as $\tau$, measures the time step at which the system reaches its absorbing state, i.e. when $\rho_X^l=1$ for the first time in the simulation.  Both the relaxation time ($\tau$) and the specific state to which the system is absorbed are random variables whose respective expected values (and higher-order moments) seem to depend on the parameters $r$ and $k$.

Furthermore, we also measure an estimator for the probability that the system will reach each of the three possible strategies as an absorbing state, as well as not reach an absorbing state at all. Here, we denote this variable as $\pi_x$ and define it as
\begin{equation}
\pi_x=\frac{1}{N_{run}}\sum_i \xi_i(x),
\label{probs}
\end{equation}
where 
\begin{equation}
\xi_i(x)=\left\{\begin{array}{cc}
1,  &\quad \mbox{if simulation reached absorbing state $x$,}  \\
0,  &  \quad \mbox{else.}
\end{array}\right.
\label{prob_counter}
\end{equation}

The estimator defined in Eq.~\ref{probs} is calculated by averaging the final result of a number $N_{run}$ of time series with different seeds each. As one might expect, $\tau\rightarrow \infty$ when the system reaches no absorbing state. To overcome this issue, we establish a maximum simulation time of $t_{max}=10^5$ after which we presume that the system does not reach an absorbing state and so we assume that $\xi=1$.




\subsection{Population size effects ($N$)}
\begin{figure}[htbp!]
\includegraphics[width=1.0\linewidth,angle=0]{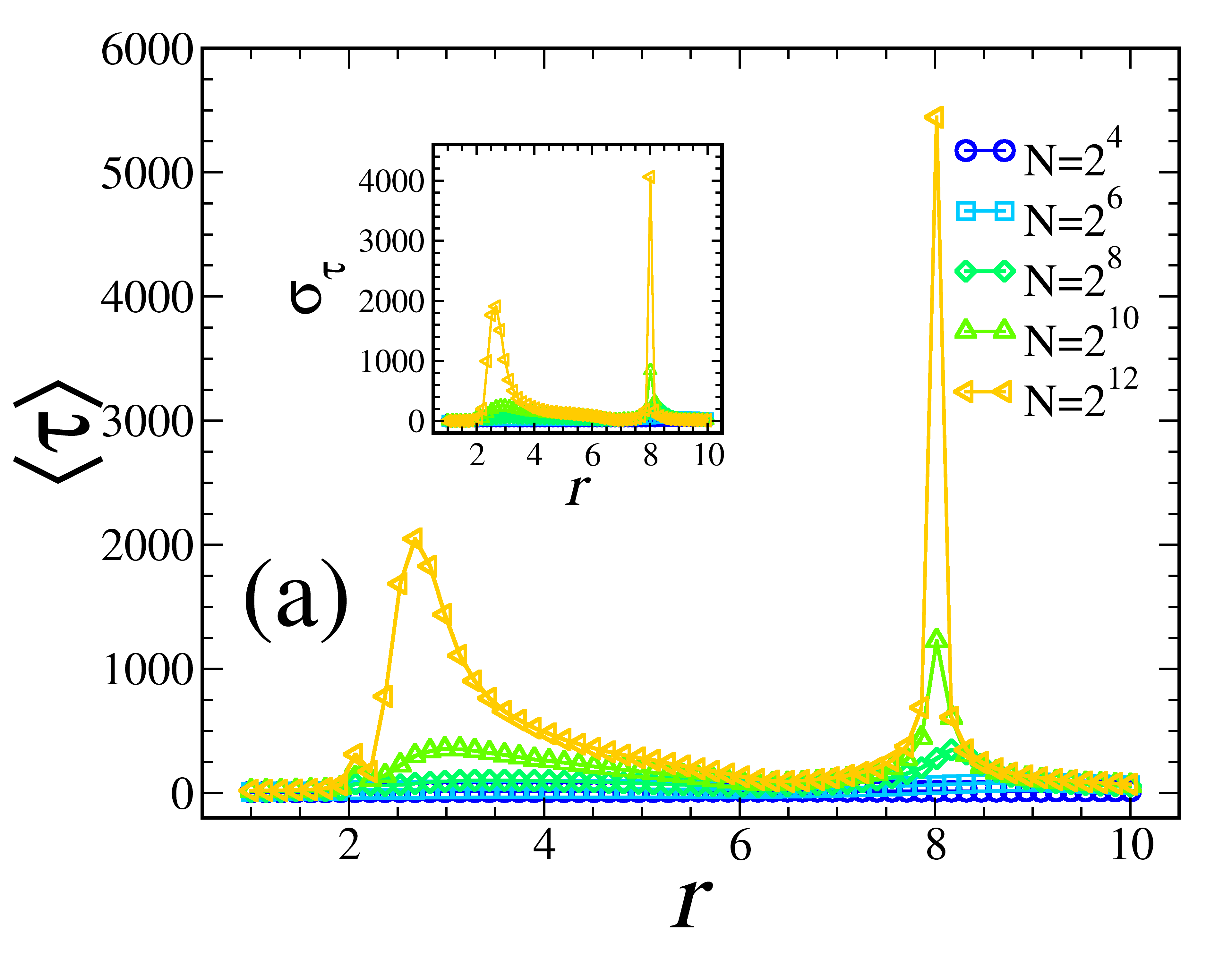}
\includegraphics[width=1.0\linewidth,angle=0]{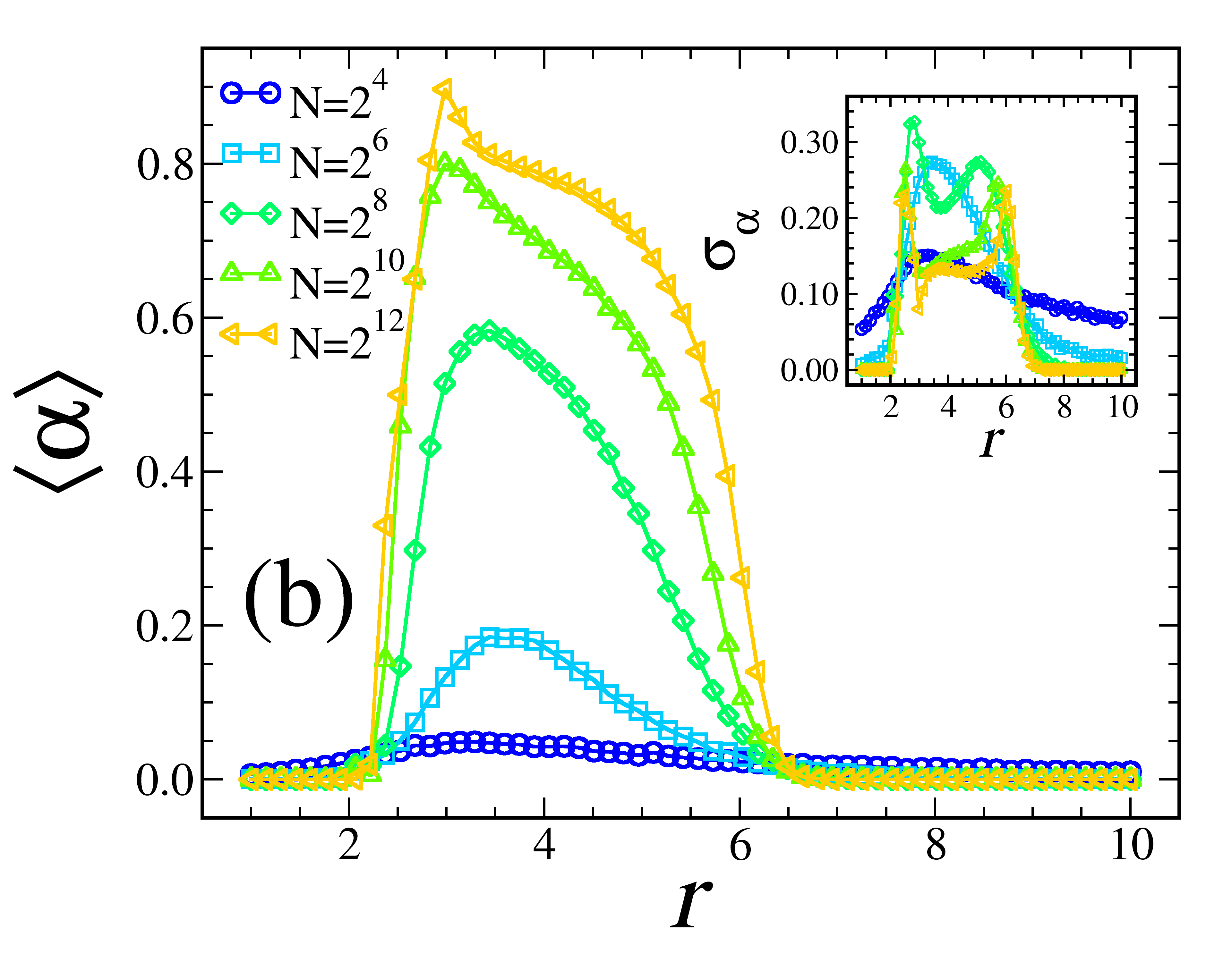}
\includegraphics[width=1.0\linewidth,angle=0]{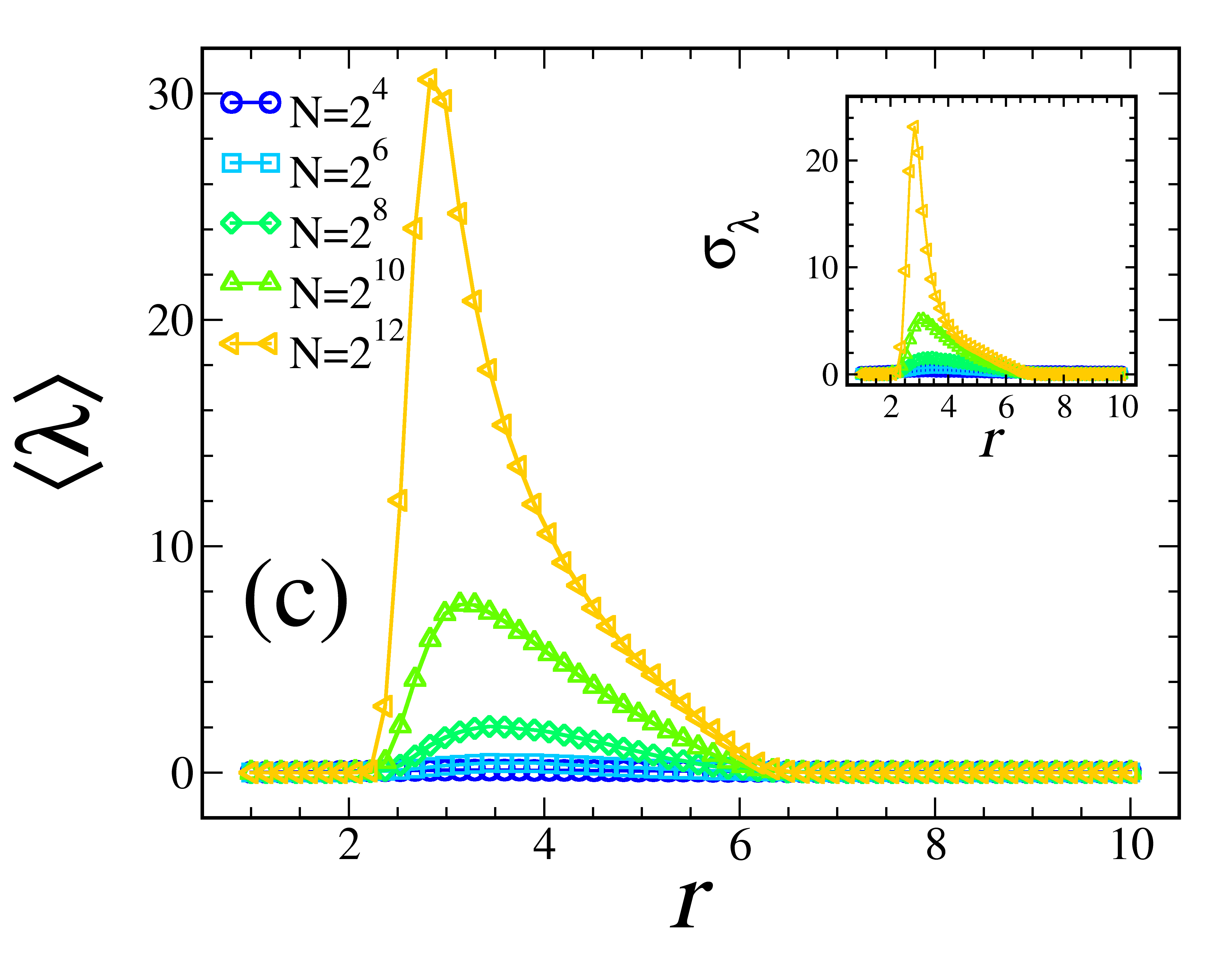}
\caption{Study of the system size considering a fixed group size $S=8$. We show the average time to reach an absorbing state (a), the average time fraction with RPS cycles (b), and the average number of cycles within the simulation time spam as functions of $r$. Each point was obtained from a sample size of $N_{run}=10^4$.}
\label{fig:population_study}
\end{figure}
\begin{figure}[htbp!]
\includegraphics[width=1.0\linewidth]{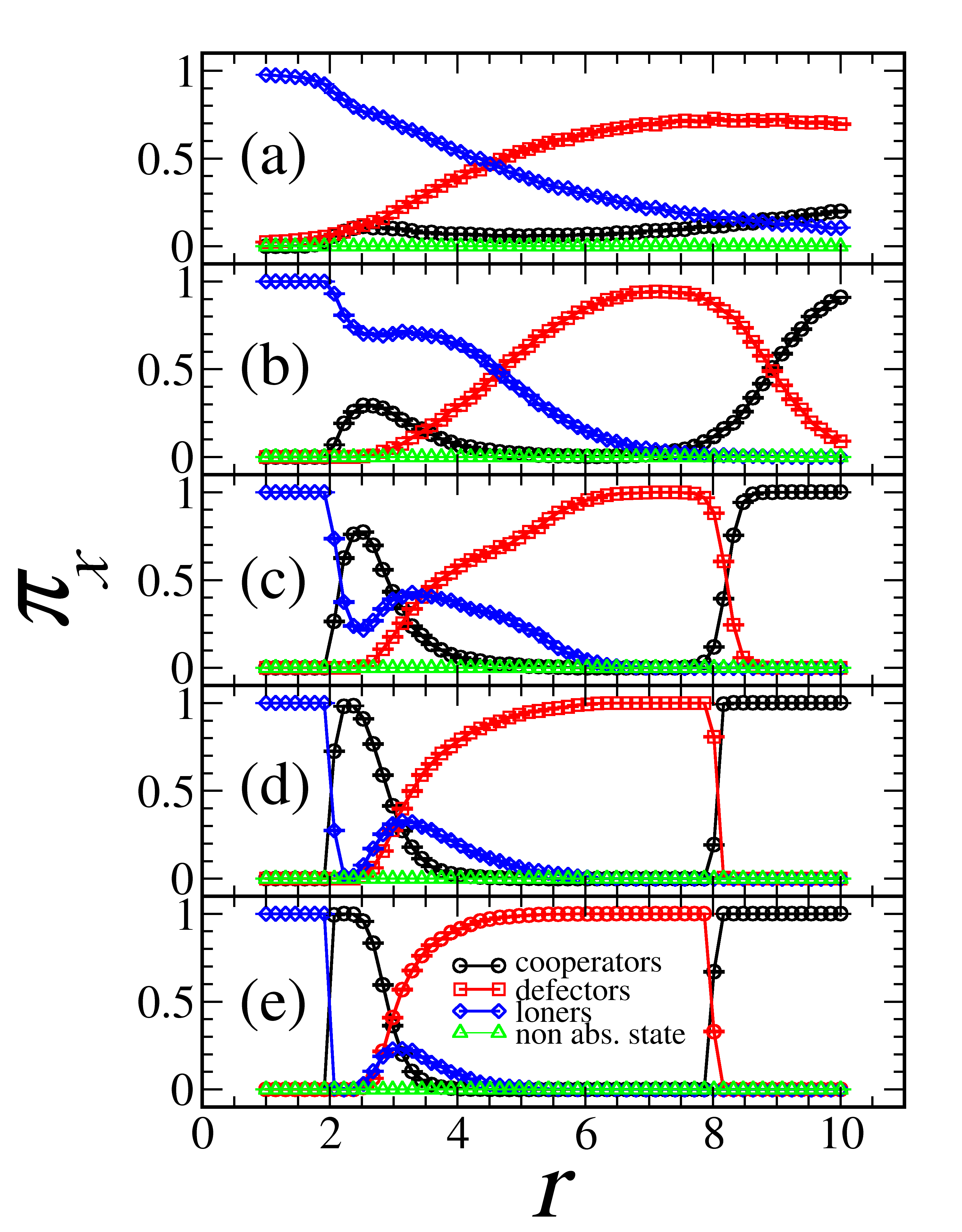}
\caption{Probabilities of reaching an absorbing state (or not) as a function of the reward $r$, for different system sizes (a) $N=2^4$, (b) $N=2^6$, (c) $N=2^8$, (d) $N=2^{10}$, and (e) $N=2^{12}$. with the group size fixed ($S=8$). Each point was obtained by averaging over $10^4$ runs.}
\label{fig:population_study_probs}
\end{figure}

A key aspect of our analysis is to determine the effects of finite-size scaling on dynamics. In the mean-field approach of Hauert et al.~\cite{Hauert2002}, the population from which groups of agents are sampled and invited to participate in a public goods game must be sufficiently large and well-mixed. Thus, for a fixed sample group size of $S=8$ and a thermal noise of $k=1$ we investigate how different population sizes ($N$) affect the relaxation time to reach an absorbing state (if it goes to one), the occurrence of RPS cycles, and the probability of reaching each possible absorbing state. We studied the expected values and standard deviation of $\tau$, $\alpha$, $\lambda$, and $\pi$ as a function of $r$ in the range $r \in [1,10]$, where each point was obtained with $N_{run}=10^4$. We fixed the maximum simulation time at $t_{\max}=10^5$ MC steps, which means that each time series that did not reach an absorbing state until this point was truncated and considered as a non-absorbing state without returning a value of $\tau$ (should be infinite theoretically) and implicating that the variable $\alpha$ is calculated over $t_{\max}$ time steps and not $\tau$ (see Eq.\ref{alpha_def} for the definition).

In Fig. \ref{fig:population_study}, we show the results for the average $\tau$ (a), the average $\alpha$ (b), and the average $\lambda$ (c). In Fig. \ref{fig:population_study_probs} we show the results for the average $\pi_X$. The first thing we notice in plots (a), (b), and (c) of Fig. \ref{fig:population_study} is that the results become more prominent as larger systems are considered. Similar things can be said about Fig. \ref{fig:population_study_probs}, with the peculiarity that, for this variable, we can identify a more clear finite-size scaling of $\pi_X$ at the ``outer boundaries'' of the region where $2 \lesssim r\lesssim 8$.

Starting from $r \approx 2$, in the inner boundaries of the region under consideration ($2 \lesssim r \lesssim 8$), we can highlight the emergence of RPS cycles for larger populations, as shown in plots (b) and (c) of Fig.~\ref{fig:population_study}. Specifically, we observe that larger systems provide a more favorable environment for the occurrence of RPS cycles. 
The peak of $\langle\alpha\rangle$ shifts slightly from $r \approx 4$ to $r \approx 3$ as the population increases to $N = 2^{12}$, as seen in Fig.~\ref{fig:population_study}(b). 
Additionally, we notice that RPS cycles have longer periods in larger systems. For instance, when comparing $N = 2^{10}$ and $N = 2^{12}$, the average period $\langle\lambda\rangle$ increases by a factor of 3, while $\langle\tau\rangle$ increases by a factor of 10 at peak. 
This, along with the increase in the RPS time fraction from $\langle\alpha\rangle \approx 0.8$ to $\langle\alpha\rangle \approx 0.9$, reinforces the conclusion that the RPS cycles 
last longer and are sustained for extended periods in larger systems.

In parallel, we note that $\langle\pi_X\rangle$ presents a more complex scenario within the region $2 \lesssim r \lesssim 8$ for $N = 2^{12}$ than Fig.~.\ref{fig:population_study} might suggest. Starting from $r \approx 2$, we observe that the system shifts its most likely absorbing state (with $\pi_L = 1$) abruptly from loner to cooperator around $r \approx 2$, maintaining this cooperative trend until $r \approx 2.5$. Beyond this point, the other two strategies gradually become more likely absorbing states, reaching a regime around $r \approx 3$ where the system is almost equally probable to relax to cooperator, loner, or defector. Interestingly, every single run in the sample reached an absorbing state for the parameters studied, as shown by the green curves in Fig.~\ref{fig:population_study_probs}, suggesting that the level of stochasticity in the strategy-switching dynamics ($k$) was high enough to prevent the system from reaching a steady coexistence state.

Finally, when $r\approx 8$, in Fig.~\ref{fig:population_study_probs}, we show that the system presents a phase transition where it abruptly switches from the known defector's economic stalemate to a scenario where the public goods game return is sufficiently high to allow the cooperation to be a more profitable strategy even if freelancer is still a valid strategy. This phase transition is also observed as the average relaxation time ($\langle\tau\rangle$) shows a peak for $r\approx 8$ as we can see in Fig. \ref{fig:population_study}(a). The precise explanation of why this happens will be given in the next subsection, where we present our results for the study of the influence of sample group sizes ($S$).

\subsection{Sample group size effects ($S$)}\label{size}
\begin{center}
\begin{figure}[htbp!]
\includegraphics[width=1.0\linewidth,angle=0]{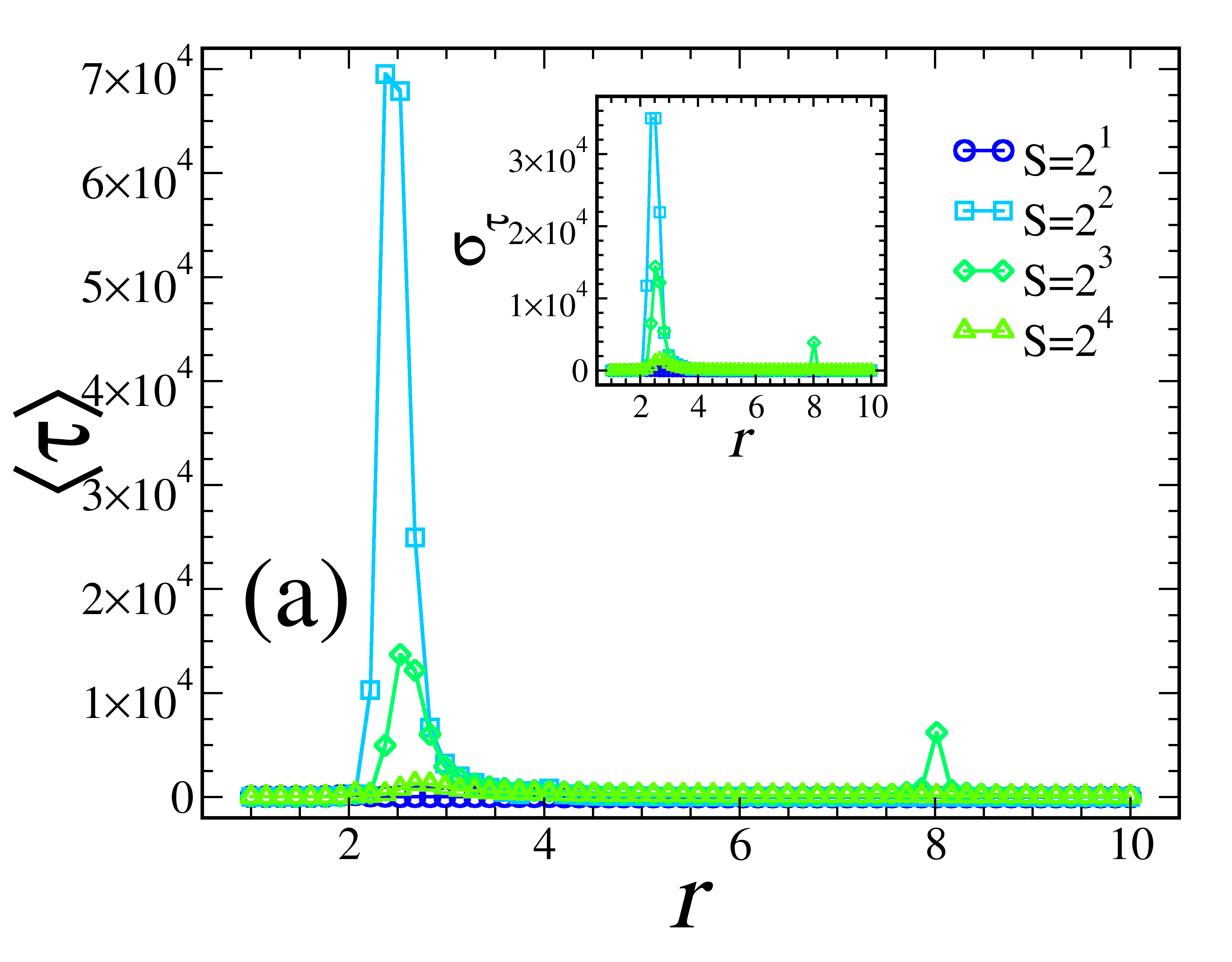}
\includegraphics[width=1.0\linewidth,angle=0]{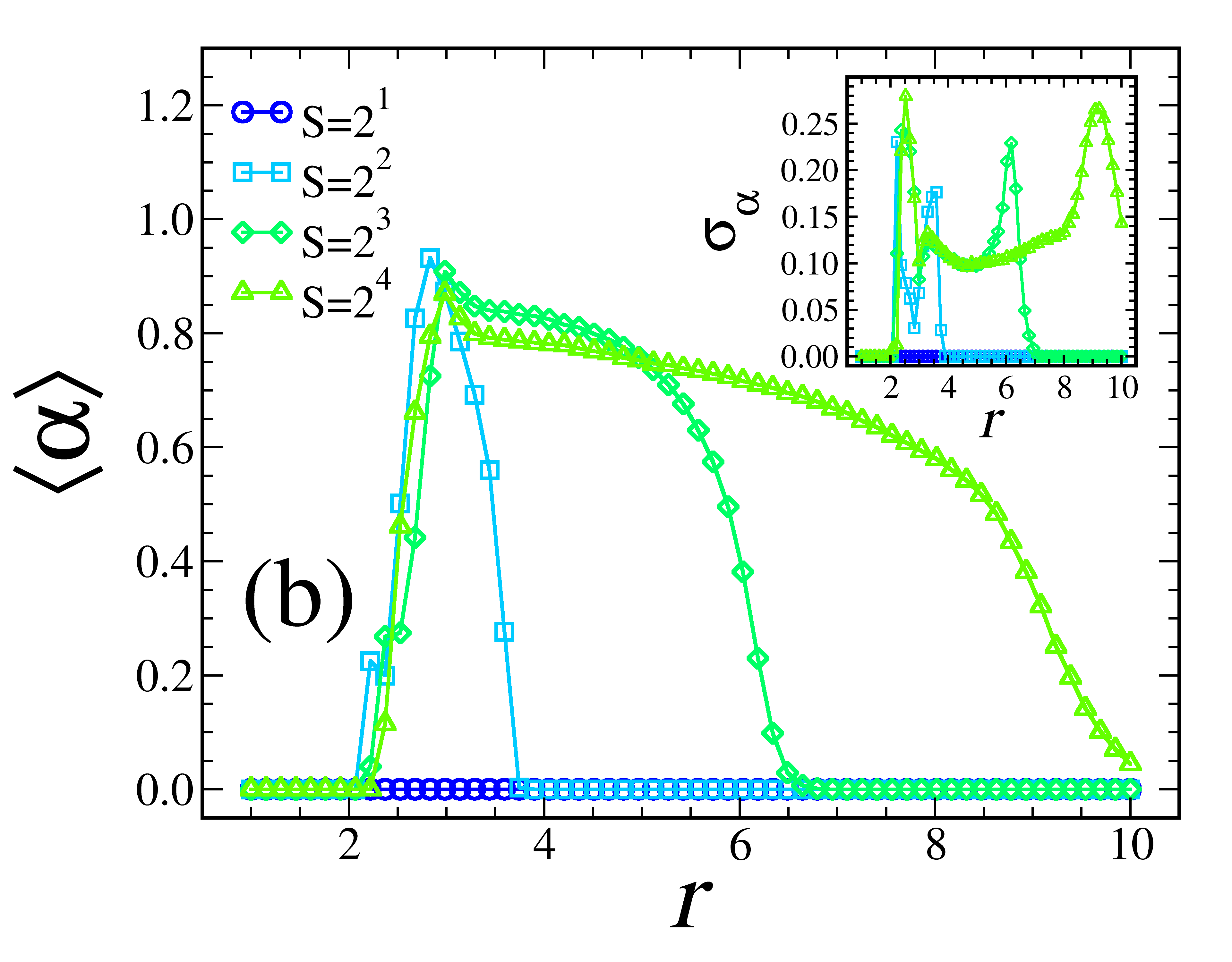}
\includegraphics[width=1.0\linewidth,angle=0]{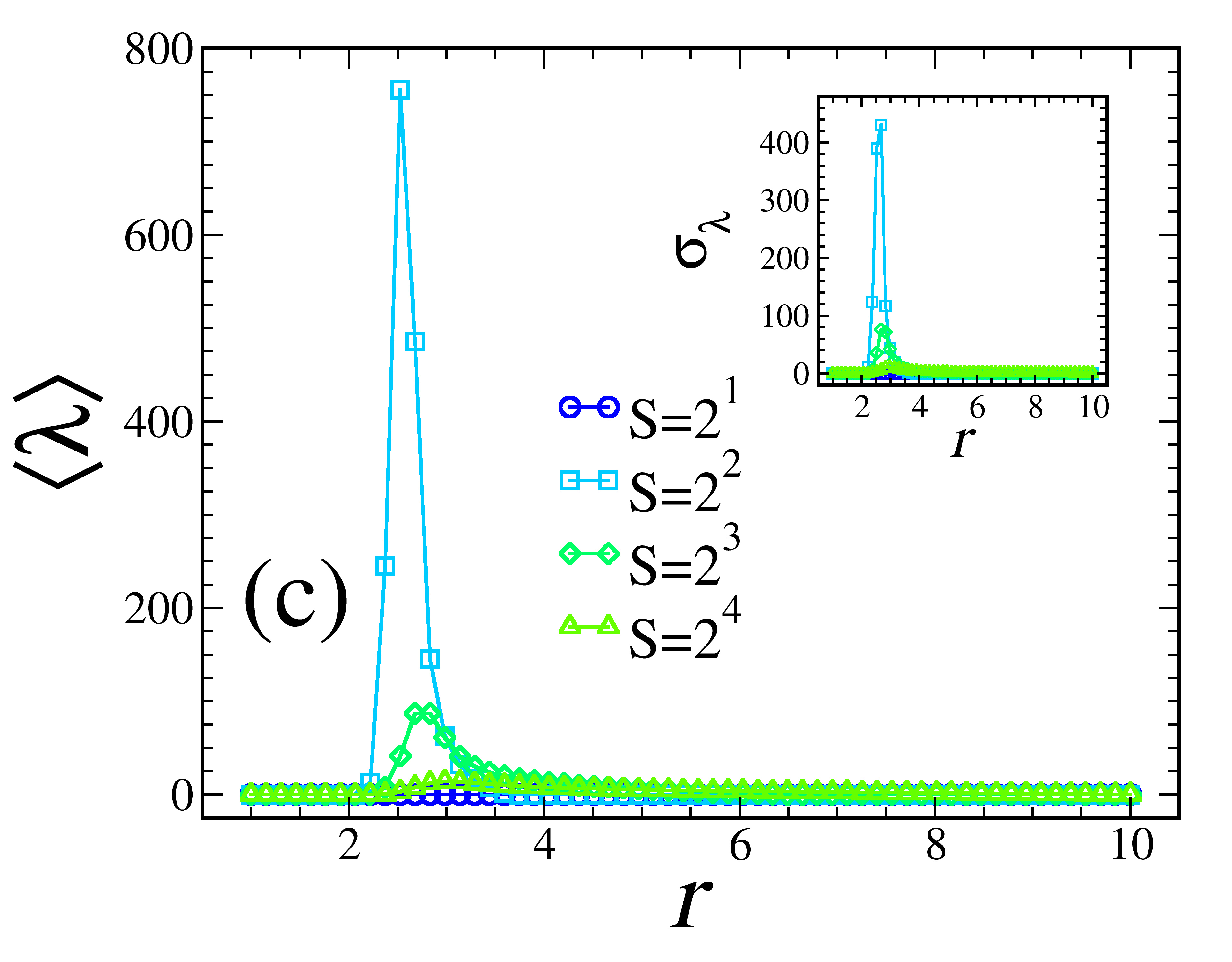}
\caption{Effect of the group size of the game on (a): the average time to reach an absorbing state, (b): the average time fraction with RPS cycles, and  (c): the average number of cycles within the simulation time span as functions of $r$. Each point was obtained in a system of size $N=8192$, averaging over $10^4$ runs.}
\label{fig:sample_group_study}
\end{figure}
\end{center}
\begin{center}
\begin{figure}[htbp!]
\includegraphics[width=1.0\linewidth]{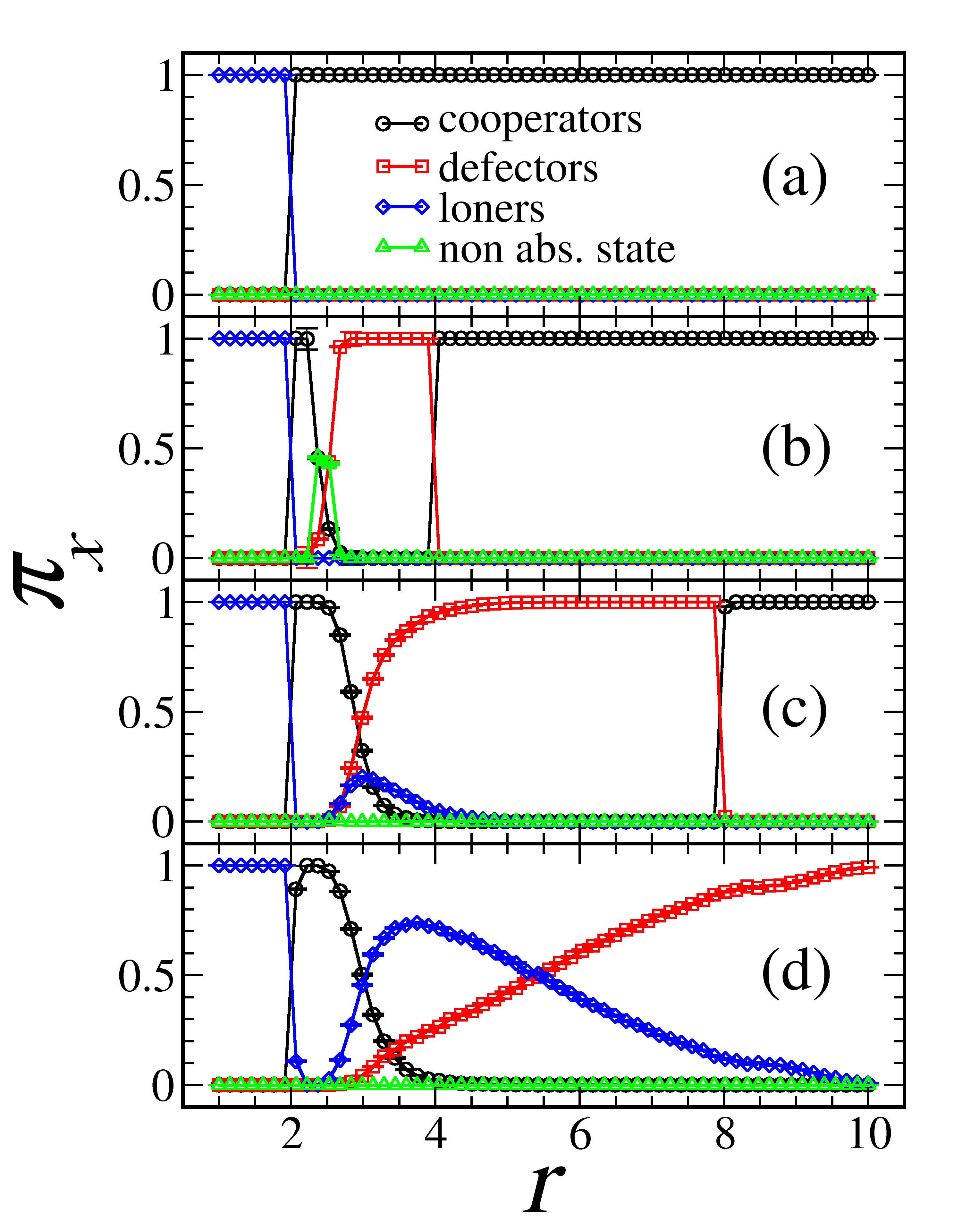}
\caption{Probabilities of reaching an absorbing state (or not) as functions of $r$ for (a) $S=2$, (b) $S=2^2$, (c) $S=2^3$, and (d) $S=2^4$. Each point was obtained for a system of size $N=8192$, averaging over $10^4$ runs.}
\label{fig:sample_group_study_probs}
\end{figure}
\end{center}

We now present the results that demonstrate how the sample group size affects the emergence of RPS cycles and the overall system dynamics. In this analysis, we fixed the population at $N = 2^{13}$ and the noise parameter at $k = 1$, and studied how the measurements $\langle\tau\rangle$, $\langle\alpha\rangle$, $\langle\lambda\rangle$, and $\pi_X$ change as a function of $r$ for different sample group sizes. 
In Fig.~\ref{fig:sample_group_study}(a), we show the average time to reach the absorbing state. We observe that the system transitions from a regime with short relaxation times for $S = 2$, where the relaxation time does not depend on the OPGG return $r$, to a sharp increase beginning around $r \approx 2$ for $S = 2^2$. For this group size, at $r \lesssim 3$, the system reaches a maximum average time of approximately $25$K MC steps to reach an absorbing state. Interestingly, for larger sample groups, such as $S \ge 2^3$, $\langle\tau\rangle$ decreases compared to $S = 2^2$, indicating that some mechanism in smaller groups is responsible for sustaining the dynamics for much longer.

The longer duration of the evolution towards an absorbing state occurs because the RPS cycles are more likely to occur with smaller groups, as we can see in Fig.~\ref{fig:sample_group_study} (b), where we show the average fraction time with RPS cycles. We observe that in the same manner that $\langle\tau\rangle$ shows a spike for $S=2^2$, $\langle\alpha\rangle$ also has a maximum value for the same group size. However, for larger group sizes such as $S=2^3$ and $S=2^4$, we observe that the RPS cycles are kept for higher values of $r$, even if it is for shorter times, as shown in Fig. \ref{fig:sample_group_study} (a). As a complementary result, we show in Fig. \ref{fig:sample_group_study} (c), that the average number of RPS cycles also shows rapid growth in the interval $2\lesssim r \lesssim 3$ when considering sample groups of size $S=2^4$.

Finally, in Fig.~\ref{fig:sample_group_study_probs}, we present the probabilities of the system reaching or not reaching an absorbing state, that is, $\pi_X$, as a function of $r$ for sample group sizes (a) $S=2$, (b) $S=2^2$, (c) $S=2^3$, and (d) $S=2^4$ using the same set of parameters as in the previous subsection. As an overview, the system shows a clear behavioral pattern that can be divided into three distinct regions. For values of $r\leq2$, the system consistently settles into an absorbing state dominated by loners. When $r\geq S$, the system relaxes into a stable state of cooperators. 
In the intermediate range, where $2 < r < S$, changes in the likelihood of the system reaching a specific steady state become more abrupt for smaller sample sizes ($S$). Specifically, for $S = 2$, the system exhibits an abrupt transition in the absorbing strategy from loner to cooperator without any intermediate metastable state.
For $S \geq 2^2$, there is a narrow region where cooperators remain the most likely absorbing state for $2 < r < 2.5$. However, near $r \approx 2.5$, the average relaxation time increases to approximately $\langle\tau\rangle = 7 \cdot 10^4$ MC steps, as shown in Fig.~\ref{fig:sample_group_study}(a), with nearly half of the runs taking even longer, {\em i.e.} $\pi_{NAS} \approx 0.5$.
When $2.5 \gtrsim r$, defectors become the most likely absorbing state, coinciding with the emergence of RPS cycles, which persist until $r \approx 3.7$. Interestingly, defectors retain their role as the most likely absorbing strategy, but only as a metastable state up to $r \leq S$.

However, for $S \geq 2^3$ (see plots (c) and (d) of Fig.~\ref{fig:sample_group_study_probs}), defectors transition to a metastable state as loners emerge as a compelling strategy. In this wider region of $r$, the three strategies are likely to occur as absorbing states.

Hauert et al.~\cite{Hauert2002} have previously pointed to the fact that cooperation prevails when $r \geq S$, however, no detailed explanation of this phenomenon has been provided. To understand the scenario in which voluntary cooperation becomes the prevailing strategy when the return ($r$) is equal to the sampled group size ($S$), let us consider two groups, $A$ and $B$, each of size $S$. Group $A$ consists of $n$ defectors and $S-n$ cooperators, while group $B$ is entirely made up of cooperators. 
So, the payoff for a defector in group $A$ is:
$$\frac{(S-n)r}{S}\;,$$ 
while the payoff of cooperators in $B$ is $r-1$.
If a defector in the group $A$ compares its payoff with any participant in the group $B$, it will shift to cooperation when the relation $$\frac{(S-n)r}{S} <  r-1\;,$$ 
is satisfied. That is, if $nr > S$.
So, the critical situation for any number of defectors $n$ occurs when $r=S$.

\section{Conclusion}\label{conclu}

We have conducted a detailed study of the conditions under which cooperation emerges in an Optional Public Good Game (OPGG), demonstrating the emergence of cyclical dynamics akin to rock-paper-scissors interactions among the three possible strategies of the system.

Our analysis reveals non-linear, oscillatory behavior in the dynamics of cooperation, where each strategy dominates in turn within certain regions of key parameters, rather than the system settling into a fixed point. These transient cycles suggest that the conditions for the emergence and stabilization of cooperation are complex, with significant implications for understanding collective action in biological, social, and economic contexts.

Previous studies have attributed the emergence of cycles in OPGG to the inclusion of the third strategy, opting out of the game (loners), which introduces cyclic dominance. However, our findings demonstrate that while the inclusion of loners is necessary for this behavior, it is not sufficient on its own.
For the spontaneous emergence of cycles, several non-trivial conditions must be met, which we summarize as follows:

\begin{itemize}

\item There are specific values of the parameter $r$ that trigger rock-paper-scissors dynamics.

\item The group size $S$ of participating players must be significantly smaller than the total population size ($S \ll N$).

\item The composition of each group changes at every iteration, ensuring no fixed interaction pattern. This setup mirrors the conditions of the social experiment reported in \cite{Fischbacher2010}.

\item The absence of a local interaction structure, {\em i.e.}, the possibility for any agent to interact with any other within a fixed group size, makes the setup analogous to a mean-field approach, leading to the observed oscillations in strategy dominance.
\end{itemize}

It should be noted that additional mechanisms are required to induce RPS cycles. In \cite{VALVERDE201761}, the authors demonstrate that dilution and mobility in two-dimensional square lattices can also trigger this phenomenon. As a result, PGG, and even more so OPGG, are compelling protocols in game theory, and their theoretical investigations and applications warrant further exploration.

\bibliography{Bibliografia}

\end{document}